\documentclass{amsproc}
\usepackage[geometry]{ifsym}
\usepackage{hyperref}
\hypersetup{
    colorlinks=true,
    linkcolor=black,
    filecolor=black,      
    urlcolor=black,
}

\newtheorem{theorem}{Theorem}[section]
\newtheorem{lemma}[theorem]{Lemma}

\theoremstyle{definition}
\newtheorem{definition}[theorem]{Definition}

\theoremstyle{remark}

\newcommand{\ds}{\displaystyle}

\numberwithin{equation}{section}

\begin{document}

\title{Preference cycles in stable matchings}


\author{Andrei Ciupan}
\thanks{Harvard University. aciupan@hbs.edu}



\date{}

\begin{abstract}

Consider the stable matching problem on two sets. We introduce the concept of a preference cycle and show how its natural presence in stable matchings proves a series of classical results in an elementary way.
\end{abstract}

\maketitle

\section{Introduction}

The problem of stable matchings has been studied by economists and applied mathematicians since Gale and Shalpey's original paper in 1962. This paper introduces a structure linking agent preferences and stable matchings. The properties of this structure provide quick and elementary proofs of many foundational results in matching theory. We first introduce the concept of stable matchings, then cary on to proving the results. Proofs of the relevant lemmas are in the appendix. We start with one-to-one matchings and then move on to many-to-one matchings.

Consider two disjoint sets $\ds M$ and $\ds W$ where each agent $\ds a$ in one of the two sets has a preference ranking of the agents of the other set, such that the ranking is strict (i.e. there are no ties) and agent $\ds a$ prefers agent $\ds b$ to agent $\ds c$ if and only if agent $\ds a$'s ranking of $\ds b$'s ranking is higher its ranking of $\ds c$. Every agent also ranks the outside option which we label $\ds \emptyset$.  This preference ranking induces a preference relation $\ds \succ_a$ for every agent. We say that $\ds a$ \emph{likes} $\ds b$ if and only if $\ds b$ is prefered to the outside option $\ds \emptyset$, i.e. $\ds b\succ_a \emptyset$.

A one-to-one matching on $\ds M\cup W$ is a function $\ds \mu : M\cup W\mapsto M\cup W \cup \{\emptyset\}$ such that $\ds \mu(m) \notin M$ if $\ds m\in M$ and $\ds \mu(W)\notin W$ if $\ds w\in W$, and $\ds \mu(\mu(a)) = a$ whenever $\ds \mu(a) \neq \emptyset$. This simply means that we can consider $\ds \mu$ as an assignment of pairs $\ds (m, \mu(m))$ of between agents in $\ds M$ and $\ds W$, with the option that some agents are unmatched, i.e. their assignment through $\ds \mu$ is the outside option $\ds \emptyset$.

A one-to-one matching is individually rational if every agent $\ds a$ likes its assignment whenever it's assigned one, i.e. $\ds \mu(a) \succ_a \emptyset$ whenever $\ds \mu(a) \neq \emptyset$.

Second, a pair $\ds (m, w) \in M\times W$ is a \emph{blocking pair} of $\ds \mu$ if $\ds m$ prefers $\ds w$ to $\ds \mu(m)$ and $\ds w$ prefers $\ds m$ to $\ds \mu(w)$. Coloquially we can interpret this as both $\ds m$ and $\ds w$ prefering to be matched to each other than staying in the assignment defined by $\ds \mu$.

Finally, a one-to-one matching $\ds \mu$ is \emph{stable} if it is individually rational and has no blocking pair.

We now define stable matching in the many-to-one case and then proceed with the results.

Consider two sets $\ds S$ and $\ds C$. It convenient to consider $\ds S$ to be a set of students and $\ds C$ a set of colleges, for simplicity and in order to quickly understand the notation that follows. Each student $\ds s\in S$ has a preference ranking over colleges in $\ds C$, and the outside option, like in the one-to-one case. For college preferences we add a few natural changed from the one-to-one case. Every college $\ds c\in C$ has a \emph{capacity} $\ds q_c$ and has strict preferences over subsets of students in $\ds S$ and the outside option again labeled $\ds \emptyset$. Finally every college $\ds c\in C$ has the property that it prefers $\ds \{c_1\}$ to $\ds \{c_2\}$ if and only if it prefers $\ds A\cup \{c_1\}$ to $\ds A\cup \{c_2\}$ for all $\ds c_1, c_2\in C, A\subset C, c_1\neq c_2, c_1, c_2 \neq A$. Such preferences are called \emph{responsive preferences.}

A many-to-one matching $\ds \mu$ on $\ds S\cup C$ is a function $\ds \mu: S\cup C \mapsto C\cup 2^S \cup \{\emptyset\}\footnote{Here $\ds 2^S$ represents the set of subsets of $\ds S$}$ such that $\ds \mu(s) \in C\cup \{\emptyset\}$ for all $\ds s\in S$ and $\ds \mu(c) = \{s\in S \mbox{ such that } \mu(s) = c\}$ for all $\ds c\in C$. These properties simply guarantee the consistence of the assignment of students to colleges, and it permits that multiple students be assigned to the same college.

A many-to-one matching $\ds \mu$ on $\ds S\cup C$ is individually rational if every student matched to a college prefers that assignment to the outside option, every college prefers each of its assigned students to the outside options, and for every college $\ds c\in C$, the number of students it is assigned, $\ds |\mu(c)$ is less than its capacity $\ds q_c$.

A pair $\ds (s, c) \in S\times C$ is a \emph{blocking pair} of $\ds \mu$ if $\ds s$ prefers $\ds c$ to $\ds \mu(c)$ and either $\ds c$ prefers $\ds s$ to one of its assigned students $\ds \mu(c)$, or $\ds c$ likes $\ds s$ and $\ds |\mu(c)| < q_c$. We can interpret this as both parties being better off if they get matched. 

Finally, a many-to-one matching $\ds \mu$ on $\ds S\cup C$ is a stable matching if it is individually rational and there are no blocking pairs. We are now ready to proceed with results.

\section{One-to-one matching}

We consider two disjoint sets $\ds M$ and $\ds W$ with strict preferences as described in the introduction. 

The deferred acceptance algorithm, the lone wolf theorem and the lattice structure of stable matchings and some of the fundamental results in matching theory. We prove the former two of these results, along with a classical result by Roth.

The following definitions will be useful:

\begin{definition} A preference cycle is an ordered list $\ds m_0w_0m_1w_1 \dots m_kw_km_0$ of unique agents, alternating between $\ds M$ and $\ds W$, such that each agent prefers its succesor in the cycle to its predecessor.
\end{definition}

\begin{definition} Consider a matching $\ds \mu$. A preference cycle $\ds m_0w_0m_1w_1 \dots m_kw_km_0$ is called $\ds (S,\mu)$ - dominating if for all agents $\ds a\in S$ in the cycle, $\ds a$'s successor in the cycle is $\ds \mu(a)$. A preference cycle is called $\ds (S, \mu) $ - dominated if for all agents $\ds a\in S$ in the cycle, $\ds a$'s predecessor in the cycle is $\ds \mu(a)$.

\end{definition}

Note that by construction, if a preference cycle is $\ds (M, \mu)$ dominating then it is also $\ds (W, \mu)$ dominated, and vice-versa.

We move on to our first result:

\begin{lemma}\label{th: 11cycle} Let $\ds \mu$ and $\mu'$ be two stable matchings on $\ds M\cup W$ with $\ds \mu'(m)\succ_m \mu(m)$ for some $\ds m\in M$. Then $\ds m$ is part of a preference cycle $\ds m_0w_0m_1w_1 \dots m_kw_km_0$ which is $\ds (M, \mu)$ dominated and $\ds (M, \mu')$ dominating.
\end{lemma}

\emph{Proof.} The proof is in the appendix.

\begin{theorem} (McVitie-Wilson '70) Let $\ds \mu, \mu'$ be two stable matchings on $\ds M\cup W$. Then $\ds \mu(a) = \emptyset \iff \mu'(a) = \emptyset$, for all $\ds a\in M\cup W$.
\end{theorem}

\emph{Proof.} Assume that there exists an agent $\ds m\in M$ for which $\ds \mu(m) = \emptyset$ and $\ds \mu'(m) \neq 0$. Then $\ds \mu'(m)\succ_m \mu(m)$. Therefore, according to Lemma \ref{th: 11cycle}, $\ds m$ is part of a $\ds (M, \mu)$  dominated preference cycle, so in particular $\ds \mu(m) \in W$, which constitutes a contradiction. $\ds\square$

\begin{theorem} (Conway '76) Let $\ds \mu, \mu'$ be two stable matchings on $\ds M\cup W$. Then there exists a stable matching $\ds \mu''$ which makes all agents in $\ds M$ weakly better off than in $\ds \mu$ or $\ds \mu'$, and all agents in $\ds W$ weakly worse off than in $\ds \mu$ or $\ds \mu'$.
\end{theorem}

\emph{Proof.} Consider the sets $\ds M'\subseteq M$ and $\ds W'\subseteq W$ of agents with different matches in $\ds \mu$ and $\ds \mu'$. According to Lemma \ref{th: 11cycle}, we can partition $\ds M'\cup W'$ into $\ds n$ disjoint preference cycles $\ds \left(m_0^{(i)}w_0^{(i)}m_1^{(i)}w_1^{(i)}\dots m_{k_i}^{(i)}w_{k_i}^{(i)}m_0^{(i)}\right)_{1\leq i\leq n}$ , with each cycle being either 

$\ds \mathcal f (M, \mu ) \mbox{  and } (W, \mu') \mbox { dominating } \mathcal g$ or 
$\ds \mathcal f (W, \mu ) \mbox{  and } (M, \mu') \mbox { dominating } \mathcal g$. 

Consider the matching $\ds \mu''$ defined as follows: if $\ds m \in M'$ then $\ds \mu''(m)$ is agent $\ds m$'s successor in $\ds m$'s corresponding preference cycle. Otherwise $\ds \mu''(m) = \mu(m) = \mu'(m)$. 

By construction, each agent in $\ds M$ is weakly better off under $\ds \mu''$ and each agent in $\ds W$ is weakly worse off under $\ds \mu''$. Let's show that $\ds \mu''$ is stable. Clearly individual rationality holds. Let's also prove that there exists no blocking pair.

Assume that agent $\ds m\in M$ prefers $\ds w \in W$ to $\ds \mu''(m)$ and that $\ds w$ prefers $\ds m$ to $\ds \mu''(w)$. Assume without loss of generality that $\ds \mu''(w) = \mu(w)$. $\ds m$ strictly prefers $\ds w$ to $\ds \mu''(m)$ and $\ds m$ weakly prefers $\ds \mu''(m)$ to $\ds \mu(m)$. Therefore $\ds (m, w)$ is a blocking pair for $\ds \mu$, which is a contradiction.

Therefore $\ds \mu''$ is indeed stable. $\ds \square$.

The following lemma will be useful for the subsequent theorem:

\begin{lemma}\label{th: 11pareto} Let $\ds \mu$ and $\ds \mu'$ be two matchings on $\ds M\cup W$ such that $\ds \mu$ is stable and $\ds \mu'(m)\succ_m \mu(m)$ for all $\ds m\in M$. Then every agent $\ds m\in M$ is part of a preference cycle that is both $\ds (M, \mu)$ dominated and $\ds (M, \mu')$ dominating. 

\end{lemma}

\emph{Proof.} The proof is in the appendix. 

\begin{theorem} (Roth '82) Let $\ds \mu$ be the stable matching on $\ds M\cup W$ obtained from the $\ds M$ - proposing deferred acceptance algorithm. Then there is no invidually rational matching $\ds\mu'$ such that $\ds \mu'(m)\succ_m \mu(m)$ for all $\ds m\in M$. 

\end{theorem}

\emph{Proof.} Assume the contrary. Then every agent $\ds m\in M$ is part of a preference cycle that is $\ds (M, \mu)$ dominated and $\ds (M, \mu')$ dominating. In particular, every agent $\ds m\in M$ has $\ds \mu(m) \in W$. Consider an agent $\ds m\in M$ which was matched with $\ds \mu(m) = w$ in the last round of the deferred acceptance algorithm. Consider $\ds m$'s corresponding preference cycle. Since $\ds w$ is $m$'s predecessor in the cycle, $\ds w$ is an element of the cycle, and also has a predecessor $\ds m_1$. $\ds w$ is $\ds m_1$'s successor in the cycle. Since this cycle is $\ds (M, \mu)$ dominated, $\ds m_1$ prefers $\ds w$ to $\ds \mu(m_1)$. Therefore $\ds m_1$ proposed to $\ds w$ at some point, and eventually got rejected. Therefore, when $\ds m$ proposed to $\ds w$, $\ds w$ was holding another agent. Since this is the last round of the algorithm, this agent is left with no match. This is a contradiction. $\ds \square$

\section{Many-to-one markets}

We consider the many-to-one matching setup on sets $\ds S \cup C$ of students and colleges, as described in the introduction. The rural hospitals theorem and Pathak and Sonmez' 08 results are two fundamental results in this space, one classical and one modern. We prove both of them with the preference cycles technique

The following definitions will be useful:

\begin{definition}A preference cycle is an ordered list $\ds s_0c_0s_1c_1\dots s_kc_ks_0$ of unique agents alternating from $\ds S$ and $\ds C$ such that each agent prefers its succesor in the cycle to its predecessor.

\end{definition}

\begin{definition} Consider a matching $\ds \mu$ on $\ds S\cup C$. A preference cycle $\ds s_0c_0s_1c_1\dots s_kc_ks_0$ of agents alternating between $\ds S$ and $\ds C$ is called $\ds (X, \mu)$ dominated if one of the following two relations holds for all agents $\ds a \in X $ in the cycle:

\begin{enumerate}
\item $\ds a\in S$ and agent $\ds a$'s predecessor in the cycle is $\ds \mu(a)$.

\item $\ds a\in C$ and $\ds a$'s predecessor in the cycle belongs to $\ds \mu(a)$.

\end{enumerate}

The analogous definition characterizes $\ds (X, \mu)$ dominating preference cycles.

\end{definition}

Note that a preference cycle is $\ds (S, \mu)$ dominating if and only if it is $\ds (C, \mu)$ dominated, and vice-versa.

Let's move on to our first result of this section:

\begin{lemma}\label{le: 1mcycle} Let $\ds \mu$ and $\ds \mu'$ be two stable matchings on $\ds S\cup C$, with $\ds \mu'(s)\succ_s \mu(s)$ for some $\ds s\in S$. Then $\ds s$ is part of a preference cycle $\ds s_0c_0s_1c_1\dots s_kc_ks_0$ which is $\ds (S, \mu)$ dominated, $\ds (S, \mu')$ dominating and all colleges in this cycle are filled at full capacity in $\ds \mu$ and $\ds \mu'$.

\end{lemma}

\emph{Proof.} The proof is in the appendix

\begin{theorem} (Rural Hospitals Theorem) Let $\ds \mu , \mu'$ be two stable matchings in a many-to-one market $\ds S\cup C$. Then the following hold:

\begin{enumerate}

\item For all $\ds s\in S$, $\ds \mu(s) = \emptyset \iff \mu'(s) = \emptyset$.

\item For all $\ds c\in C$, $\ds |\mu(c)| = |\mu'(c)|$.

\item If $\ds c\in C$ and $\ds \mu(c) \neq q_c$ (the college's capacity), then $\ds \mu(c) = \mu'(c)$.

\end{enumerate}

\end{theorem}

\emph{Proof.} We prove the results in order. First, assume without loss of generality that $\ds \mu'(s) \succ_s \mu(s) = \emptyset$. We apply lemma \ref{le: 1mcycle} and get that $\ds s$ is a part of a preference cycle which is $\ds (S, \mu)$ dominated and $\ds (S, \mu')$ dominating. This means that $\ds \mu(s)$ is $\ds s$'s predecessor in the cycle, which is an element of $\ds C$, which is a contradiction. $\ds \square$

For the second result, assume that $\ds |\mu(c)| > |\mu'(c)|$ for some college $\ds c$. Therefore there exists a student $\ds s$ such that $\ds \mu(s) = c$ and $\ds \mu'(s) \neq c$. Since $\ds c$ has responsive preferences, $\ds c$ would prefer to include $\ds s$ to its matched set $\ds \mu'(s)$. Since $\ds \mu'$ is stable, we must have $\ds \mu'(s) \succ_s c = \mu(s) $. Therefore $\ds s$ and $\ds c$ are part of a preference cycle which is $\ds (S, \mu')$ dominated, and $\ds c$ is filled at full capacity under $\ds \mu'$ and $\ds \mu'$, which is a contradiction. $\ds \square$.

For the third result, assume that $\ds \mu(c) \neq \mu'(c)$ for some college $\ds c$. Let $\ds s$ be the highest-ranked student from $\ds \mu(c) \cup \mu'(c)$ which is matched to $\ds c$ in one of the two matchings $\ds \mu, \mu'$ but not in both. Assume without loss of generality that $\ds \mu(s) = c$. Since preferences are responsive, in stable matching $\ds \mu'$, $\ds c$ would prefer to include $\ds s$, either by replacing its least prefered student in $\ds \mu'(c)$ or by adding $\ds s$ to the class $\ds \mu'(c)$ if there is capacity. Since $\ds \mu'$ is stable, $\ds s$ prefers $\ds \mu'(s)$ to $\ds c$. This means that $\ds s$ and $\ds c$ are part of a $\ds (S, \mu')$ dominating preference cycle and $\ds c$ is at full capacity in matching $\ds \mu'$, which implies that $\ds c$ is at full capacity in matching $\ds \mu$, which constitutes a contradiction. $\ds \square$.

\begin{theorem} (Pathak-Sonmez '08). Consider the Boston mechanism on $\ds S\cup C$, where $\ds S$ is partitioned into two sets, $\ds S = N \sqcup M$, such that the action set of players in $\ds N$ only contains the truthful strategy and the action set of players in $\ds M$ contains all strategies of this game, i.e. players in $\ds N$ are sincere and players in $\ds M$ are sophisticated. Then the college allocation of any sincere student is the same across all Nash Equilibria of this game.

\end{theorem}

\emph{Proof.} First of all, we know from Pathak - Sonmez '13 that every Nash Equibrium of the Boston mechanism is equivalent to a stable matching in the economy where students have the same preferences as before, and college preferences are characterized by tiers : all students in a higher tier are preferred to students in a lower tier, students in the same tier are ranked according to the original preferences of the college, the first tier contains all sophisticated students along with sincere students who ranked that college as number one, and tier $\ds k\neq 1$ contains all sophisticated students who ranked that college as number $\ds k$.

Assume that a sincere student $\ds s$ gets two different college assignments in different Nash Equilibria. From now on, focus on the preferences of the modified economy. $\ds s$ has two different college assignments in two stable matchings. Assume these matchings are $\ds \mu$ and $\ds \mu'$ and assume $\ds \mu'(s) \succ_s \mu(s)$. From lemma \ref{le: 1mcycle}, $\ds s$ is part of a preference cycle that is $\ds (S, \mu)$ dominated and $\ds (S, \mu')$ dominating. Since $\ds s$ is sincere and $\ds s$ prefers its successor $\ds c_1$ in the cycle to its predecessor $\ds c$, college $\ds c$ does not have $\ds s$ in its first tier. Since $\ds c$ nevertheless prefers $\ds s$ to $\ds c$'s predecessor $\ds s_0$, it must be the case that $\ds s_0$ is in a tier at most as high as $\ds s$ under $\ds c$'s preferences. Therefore $\ds s_0$ is also sincere.

Recall that each student's successor in the preference cycle is its matching under $\ds \mu'$ and its predecessor is its matching under $\ds \mu$. Consider matching $\ds \mu$. $\ds s$ is matched to $\ds c$, but $\ds s_0$ is not matched with $\ds c$ but with $\ds s_0$'s predecessor in the preference cycle, $\ds c_0$. We know that $\ds s_0$ prefers $\ds c$ to $\ds c_0$, therefore since $\ds s_0$ is sincere, $\ds s_0$ applied to $\ds c$ first, and got rejected. Since $\ds s$ got matched to $\ds c$, it must be the case that $\ds s$ applied to $\ds c$ at least as early as $\ds s_0$ applied to $\ds c$, which means that $\ds s$ got matched in $\ds \mu$ strictly earlier than $\ds s_0$ got matched in $\ds \mu$.

We just showed that if a student $\ds s$ in this preference cycle is sincere, then its predecessor's predecessor, $\ds s_0$, is sincere and matched in $\ds \mu$ strictly after $\ds s$. Iterating this process, since we are iterating through predecessors in a cycle, we will get back to $\ds s$ and reach a contradiction: $\ds s$ being matched in $\ds \mu$ after itself. Therefore the initial assumption was false. $\ds \square$.

\section{Appendix}

Here we prove the three lemmas from the paper.

\textbf{Lemma \ref{th: 11cycle}.} Let $\ds \mu$ and $\mu'$ be two stable matchings on $\ds M\cup W$ with $\ds \mu'(m)\succ_m \mu(m)$ for some $\ds m\in M$. Then $\ds m$ is part of a preference cycle $\ds m_0w_0m_1w_1 \dots m_kw_km_0$ which is $\ds (M, \mu)$ dominated and $\ds (M, \mu')$ dominating.

\emph{Proof.} Denote $\ds m_0 = m$. Since $\ds \mu'(m_0)\succ_{m_0} \mu(m_0)$, it must be the case that $\ds \mu'(m_0) \in W$, and let $\ds \mu'(m_0) = w_0$. In matching $\ds \mu$, $\ds m_0$ prefers $\ds w_0$, so it must be the case that $\ds w_0$ is matched to an agent $\ds m_1 \in M, m_1\neq m_0$, such that $\ds w_0$ prefers $\ds m_1$ to $\ds m_0$. But then, in matching $\ds \mu',$ it must be the case that $\ds m_1$ is matched to an agent $\ds w_1\in W$, $\ds w_1 \neq w_0$, such that $\ds m_1$ prefers $\ds w_1$ to $\ds w_0$. Now, consider that we have continued this process, and take the largest set of distinct agents $\ds m_0, m_1, \dots m_k \in M$ and $\ds w_0, w_1, \dots w_k$ such that the following conditions all hold:

\begin{enumerate}

\item $\ds \mu(m_i) = w_{i-1}$ and $\ds \mu'(m_i) = w_i$ for $\ds 1\leq i\leq k$.

\item $\ds m_i$ prefers its matching in $\ds \mu'$ to its matching in $\ds \mu$, for $\ds 0\leq i\leq k$.

\item $\ds w_i$ prefers its matching in $\ds \mu$ to its matching in $\ds \mu'$, for $\ds 0\leq i< k$.

\end{enumerate}

Now consider agent $\ds m_k$ in matching $\ds \mu$: it prefers $\ds w_k$ to its match $\ds w_{k-1}$. Therefore, in $\ds\mu$, agent $\ds w_k$ has a match which it prefers to $\ds m_k$.

Assume $\ds \mu(w_k) \neq m_0$. Then let $\ds \mu(w_k) = m_{k+1}$ with $\ds m_{k+1} \notin \{m_0, m_1, \dots, m_k\}$, and consider $\ds w_k$ in matching $\ds \mu'$: it prefers $\ds m_{k+1}$ to $\ds m_k$. Therefore $\ds m_{k+1}$ must be matched in $\ds \mu'$ to someone it prefers to $\ds w_k$. 

Let $\ds \mu'(m_{k+1}) = w_{k+1}$. Since $\ds m_{k+1} \notin  \{m_0, m_1, \dots, m_k\}$, we must also have $\ds \mu'(m_{k+1}) \notin  \{\mu'(m_0), \mu'(m_1), \dots, \mu'(m_k)\} =  \{w_0, w_1, \dots, w_k\}$. Now notice that the larger set of agents $\ds m_0, \dots, m_k, m_{k+1}, w_0, \dots, w_k, w_{k+1}$ also satisfies the three conditions above. This is a contradiction to our construction. Therefore $\ds \mu(w_k) = m_0$, so $\ds m_0w_0m_1w_1\dots m_kw_km_0$ is a preference cycle where for every $\ds m\in M$ in the cycle, $\ds \mu(m)$ is $\ds m$'s predecessor and $\ds \mu'(m)$ is $m$'s successor, so indeed this cycle is $\ds (M, \mu)$ dominated and $\ds (M, \mu')$ dominating, as desired. $\ds\square$

\textbf{Lemma \ref{th: 11pareto}.} Let $\ds \mu$ and $\ds \mu'$ be two matchings on $\ds M\cup W$ such that $\ds \mu$ is stable and $\ds \mu'(m)\succ_m \mu(m)$ for all $\ds m\in M$. Then every agent $\ds m\in M$ is part of a preference cycle that is both $\ds (M, \mu)$ dominated and $\ds (M, \mu')$ dominating. 

\emph{Proof.} Consider $\ds m_0 \in M$. Since $\ds \mu'(m_0) \succ_m \mu(m) \succeq_m \emptyset,$ we have $\ds \mu'(m_0) \in W$. Let $\ds \mu'(m_0) = w_0$. Since $\ds m$ prefers $\ds m_0$ to $\ds \mu(m_0)$ and $\ds \mu$ is stable, $\ds w_0$ must be matched in $\ds \mu$ to an agent $\ds m_1 \in M$ which $\ds w_0$ prefers to $\ds m$. The argument now is identical to the one in the proof of the previous lemma: Consider the largest path $\ds m_0 w_0 m_1w_1\dots m_kw_km_{k+1}$ of distinct agents alternating between $\ds M$ and $\ds W$ with the property that each agent $\ds w_i$'s predecessor is $\ds \mu'(w_i)$ and its successor is $\ds \mu(w_i)$, for $\ds 0\leq i\leq k$ and each agent prefers its successor to its predecessor. 

Since $\ds \mu'(m_{k+1}) \succ_{m_{k+1}} w_k$, let $\ds \mu'(m_{k+1}) = w_{k+1} \in W$. Since $\ds m_{k+1} \notin \{m_0, m_1, \dots m_k\}$, we have $\ds w_{k+1} = \mu'(m_{k+1}) \notin \{\mu'(m_0), \mu'(m_1), \dots \mu'(m_k)\} = \{w_0, m_1, \dots w_k\}$. In stable matching $\ds \mu$, $\ds m_{k+1}$ prefers $\ds w_{k+1}$ to $\ds w_k$. Therefore $\ds w_{k+1}$ must be matched with an agent $\ds m_{k+2}$ which it prefers to $\ds m_{k+1}$. If $\ds m_{k+2} \notin  \{m_0, m_1, \dots ,m_{k+1}\}$ then we have reached a longer path $\ds m_0 w_0 m_1w_1\dots m_kw_km_{k+1}w_{k+1}m_{k+2}$, which would be a contradiction. Therefore $\ds m_{k+2} \in \{m_0, m_1, \dots ,m_{k+1}\}$, or equivalently 

$\ds \mu(w_{k+1}) \in \{m_0, \mu(w_0), \mu(w_1), \dots ,\mu(w_k)\}$, which implies that $\ds m_{k+2} = m_0$, so indeed $\ds m_0$ is part a preference cycle which is $\ds (M, \mu)$ dominated and $\ds (M, \mu')$ dominating. $\ds \square$

\textbf{Lemma \ref{le: 1mcycle}.} Let $\ds \mu$ and $\ds \mu'$ be two stable matchings on $\ds S\cup C$, with $\ds \mu'(s)\succ_s \mu(s)$ for some $\ds s\in S$. Then $\ds s$ is part of a preference cycle $\ds s_0c_0s_1c_1\dots s_kc_ks_0$ which is $\ds (S, \mu)$ dominated, $\ds (S, \mu')$ dominating and all colleges in this cycle are filled at full capacity in $\ds \mu$ and $\ds \mu'$.

\emph{Proof.} Needs more work!

\end{document}